# Regular realizability problems and models of a generalized nondeterminism

A. Rubtsov[*]  M. Vyalyi[†]
rubtsov99@gmail.com  vyalyi@gmail.com

May 29, 2011

Models of a generalized nondeterminism are defined by limitations on nondeterministic behavior of a computing device. A regular realizability problem is a problem of verifying existence of a special sort word in a regular language. These notions are closely connected.

In this paper we consider regular realizability problems for languages consisting of all prefixes of an infinite word. These problems are related to the automata on infinite words and to the decidability of monadic second-order theories.

The main contribution is a new decidability condition for regular realizability problems and for monadic-second order theories.

We also show that decidability of a regular realizability problem is equivalent to decidability of some prefix realizability problem.

Models of a generalized nondeterminism for multi-head 2-way automata were introduced in [4, 5]. In these models a computing device has an access to an additional data (a *guess*) stored in an auxiliary memory. Generalized models are also obtained by specifying the memory structure or by restricting the guess content. In this paper we adopt the latter approach. It leads to a notion of generalized nondeterministic multi-head 2-way automaton (GNA).

It appears that many standard complexity classes can be expressed in terms of generalized nondeterminism, see [4].

In this paper we consider decidability questions only.

There is a natural complete problem for each GNA model—so called *regular realizability problem*. A particular regular realizability problem is specified by a language $L$ (a *filter*). The question is whether the filter has a non-empty intersection with a given regular language. A complete regular realizability problem for a GNA model has a specific structure. Namely, a filter consists of all prefixes of infinite words taken from possible guesses in the GNA model.

We consider here an extreme case when there is the unique guess. It means that the filter of a regular realizability problem is the language of infinite word's prefixes. We call a problem of this sort a *prefix realizability problem*. In this case a nondeterminism become degenerated and such a model can be considered as a generalization of an oracle model of computation.

[*]Moscow Institute of Physics and Technology. Supported by RFBR grant 11–01–00398
[†]Dorodnitsyn Computing Center of RAS. Supported by RFBR grants 09–01–00709, 11–01–00398

We show that decidability of a regular realizability problem is equivalent to decidability of a prefix realizability problem.

A prefix realizability problem can be represented as a question about occurrence of regular event in an infinite word. The most interesting example of problems concerning regular events in infinite words is a problem of acceptance by nondeterministic Büchi automaton. It was shown by Büchi [1] that decidability of this problem is equivalent to decidability of a specific monadic second-order theory (MSO) of natural numbers with order relation extended by unary predicates that indicate positions of a particular symbol in an infinite word.

Using this technique several conditions of decidability for MSO theories were established. The works of Semenov [2] and Carton and Thomas [3] led out two broad classes of decidable words. Semenov proved decidability for almost periodic words. Carton and Thomas proved decidability for morphic words.

Prefix realizability problems are reducible to the corresponding problems for Büchi automata. So decidability of an MSO implies decidability of the corresponding prefix realizability problem. In other direction we give an example of a prefix decidable but Büchi undecidable word.

We suggest here a simple new decidability condition for prefix realizability problems and Büchi realizability problems. An infinite word is prefix decidable (Büchi decidable) if any finite word is a factor of this infinite word.

To prove an equivalence of any regular realizability problem to some prefix realizability problem we generalize this condition to infinite alphabets.

## 1 Problems for automata reading infinite words

Let $L$ be a language over a finite alphabet $\Sigma$. We call the language $L$ a *filter*. *Regular $L$-realizability problem* is an algorithmic problem. The input of the problem is a description of a regular language $R$ over the alphabet $\Sigma$. The question is to verify that $L \cap R \neq \varnothing$. Regular $L$-realizability problem is denoted by $\mathrm{RR}(L)$.

We are interested here in decidability of regular realizability problems. In this case all forms of regular languages descriptions are equivalent. Our basic form is description by a deterministic automaton recognizing the language but we also use nondeterministic automata and regular expressions as well.

The transition function of an automaton $A$ extended to the set of all words over the input alphabet is denoted by $\delta_A(u, q)$. (For nondeterministic automata the transition function is changed by the transition relation.)

Now we introduce a specific form of a regular realizability problem.

*An infinite word* over the alphabet $\Sigma$ is a sequence $W = w_1 w_2 \ldots w_n \ldots$ of symbols from $\Sigma$. We denote by $W[n, m]$ a word $w_n \ldots w_m$, which is called a *factor*. Factors in the form $W[1, n]$ are called *prefixes*. We also refer to $w_n$ as $W[n]$.

An infinite word $W = w_1 w_1 \ldots w_n \ldots$ is *computable* if a function $n \mapsto w_n$ is computable. In the sequel we consider computable words only and omit this requirement in the statements.

Prefixes of an infinite word $W$ form a language denoted by $\mathrm{Pref}(W)$. The regular $\mathrm{Pref}(W)$-realizability problem is called *prefix realizability problem* and denoted by $\mathrm{R}_p(W)$. We call an infinite word $W$ decidable if $\mathrm{RR}(\mathrm{Pref}(W))$ is decidable.

*Büchi-realizability problem* $\mathrm{R}_p^\infty(W)$ is the question whether $|R \cap \mathrm{Pref}(W)| = \infty$. An infinite word is Büchi decidable if $\mathrm{R}_p^\infty(W)$ is decidable.

Changing in previous definitions prefixes by factors we obtain *factor realizability problem* $\mathrm{R}_f(W)$ and $\infty$-*factor realizability problem* $\mathrm{R}_f^\infty(W)$. Similarly, an infinite word is called *factor decidable* if $\mathrm{R}_f(W)$ is decidable and $\infty$-*factor decidable* if $\mathrm{R}_f^\infty(W)$ is decidable.

There are natural reductions between these problems.

**Proposition 1.** $\mathrm{R}_f(W) \leqslant_m \mathrm{R}_p(W) \leqslant_m \mathrm{R}_p^\infty(W)$ *and* $\mathrm{R}_f^\infty(W) \leqslant_m \mathrm{R}_p^\infty(W)$.

Here $\leqslant_m$ is the $m$-reducibility (the mapping reducibility).

*Proof.* Factor problems are reduced to prefix ones by mapping $R$ to $\Sigma^* R$.

The reduction $\mathrm{R}_p(W) \leqslant_m \mathrm{R}_p^\infty(W)$ maps an input automaton $A$ of the problem $\mathrm{R}_p(W)$ to an automaton $\tilde{A}$. The last one differs from $A$ in transitions from accepting states. Each accepting state of $A$ becomes absorbing in the automaton $\tilde{A}$: if $q \in F(A)$ then $\delta_{\tilde{A}}(a, q) = q$ for all $a \in \Sigma$. □

## 2 Büchi-realizability and prefix realizability

In this section we show that Büchi decidable words correspond to decidable monadic second-order theories $\mathrm{MT}(\mathbb{N}, <, W)$, $W \in \Sigma^\infty$. The monadic second-order theory is an extension of the first-order theory of integers with the order relation and unary predicates $a(n)$ for each symbol $a \in \Sigma$. Additionally, an MSO formula may contain monadic variables over unary predicates. In interpretation of the formula a predicate $a(n)$ is true if $w_n = a$ and the order relation is the standard order on the natural numbers.

Let $\phi$ be a formula of MSO. Denote by $L^\infty(\phi)$ a set of infinite words $W$ such that $\phi$ is true in $\mathrm{MT}(\mathbb{N}, <, W)$. It appears that for any formula $\phi$ the set $L^\infty(\phi)$ is an $\omega$-regular language, i.e. a set of infinite words accepted by an automaton. For infinite words different definitions of automata lead to different classes of $\omega$-regular languages.

Definitions of $\omega$-automata depend on a choice between deterministic and nondeterministic automata and on an acceptance rule.

*Run* of an automaton $A$ with the state set $Q$ on an infinite word $W$ is an infinite sequence over the alphabet $Q$

$$\rho = q_0 q_1 \ldots q_n \ldots \tag{1}$$

such that $q_{n+1}$ can be produced from $q_n$ by reading $n$th symbol of the word $W$. For deterministic automata the run is unique.

*The limit set* $\lim_\rho A$ consists of those states that appear infinitely often in the sequence (1).

A *Büchi automaton* accepts an infinite word $W$ if for some run $\rho$ on the word $W$ an accepting state belongs to the limit set $\lim_\rho A$.

In a *Muller automaton* the set of accepting states is changed by a family of accepting macrostates $\mathcal{F} \subseteq 2^Q$, where $Q$ is the state set of the automaton. The Muller automaton accepts a word if $\lim_\rho A \in \mathcal{F}$ for some run $\rho$.

By $L^\infty(A)$ we denote the set of infinite words accepted by an automaton $A$ (a $\omega$-regular language above). It appears that nondeterministic Büchi and Muller automata (either deterministic or nondeterministic) accept the same class of $\omega$-languages [7, 8]. Moreover, the constructions of equivalent automata is effective (i.e. can be performed by an algorithm).

**Theorem** (Büchi, [1]). *There exists an algorithm such that builds from a nondeterministic Büchi automaton $A$ an equivalent MSO formula $\phi$, i.e. $L^\infty(A) = L^\infty(\phi)$. Also, there exists an inversed algorithm that builds from a formula an equivalent Büchi automaton.*

Deterministic Büchi automata accept a smaller class of $\omega$-languages. But in decidability questions this difference is irrelevant.

**Proposition 2.** *Checking acceptance by a nondeterministic Büchi automaton is Turing reducible to the Büchi-realizability problem $\mathrm{R}_p^\infty(W)$.*

*Proof.* Let $N$ be a nondeterministic Büchi automaton and $M$ is an equivalent deterministic Muller automaton.

Now we construct a family of deterministic Büchi automata $D_F$, where $F \in \mathcal{F}(M)$. The states of an automaton $D_F$ are macrostates (i.e. subsets of the state set) of the Muller automaton $M$. Transitions in $D_F$ are induced by transitions in the Muller automaton
$$\delta_D(a, S) = \{q : q = \delta_M(a, q'),\ q' \in S\}.$$
The only accepting state of the $D_F$ is $F$.

By definition the automaton $M$ accepts an infinite word $W$ iff some macrostate from the family $\mathcal{F}(M)$ appears infinitely often on the automaton run. It means that for some automaton $D_F$ the answer in Büchi-realizability problem is positive. $\square$

From Proposition 1 we see that Büchi decidability implies prefix decidability. Thus from Proposition 2 and Büchi theorem we conclude that decidability of the theory $\mathrm{MT}(\mathbb{N}, <, W)$ implies prefix decidability.

So, we can directly apply known results on MSO decidability to obtain prefix decidable words. In particular, results of [2, 6] give us prefix decidability of efficiently generalized almost periodic words and the main result of [3] implies that morphic words are prefix decidable.

There is no reduction in the opposite direction.

**Theorem 1.** *There is a prefix decidable but Büchi undecidable word $W$.*

*Proof.* Let fix a computable enumeration of Turing machines and a computable bijection between finite deterministic automata and natural numbers.

The word $W \in \{0,1\}^\infty$ has the form

$$w_1 u_1 w_2 u_2 \ldots w_n u_n \ldots,$$

where finite words $w_n$ and $u_n$ are block concatenations. A *block* is a word of the following form $b_m = 10^m 1$ ($m$ is called a *block rank*).

Let $T_n$ be the numbers $k$ of Turing machines satisfying the following conditions: $k \leqslant n$ and $k$th Turing machine doesn't stop after $n$ steps running on the empty input. By definition $T_n$ is finite.

The word $w_n$ is a concatenation of blocks with block ranks $j \in T_n$ in the ascending order of $j$.

The construction of the word $u_n$ depends on behavior of the automaton $A_n$ on words of the form

$$w_1 u_1 w_2 u_2 \ldots w_n E,$$

where $E$ is a concatenation of blocks corresponding to the machines that do not stop at this stage.

More formally, define a set

$$S_n = \{b_k : (k > n) \vee (k \in T_n)\}$$

and a language $S_n^*$. Let $q$ be the state of the automaton $A_n$ occurred after reading the word $v_n$. Build an automaton $A_n'$ from the $A_n$ by changing the initial state to $q$.

If $L(A_n')\Sigma^* \cap S_n^* \neq \varnothing$ then $u_n$ is the lexicographically smallest word of this language. Otherwise, $u_n = \varepsilon$ (the empty word).

This completes the construction of the word $W$.

Prove that $W$ is computable. It is obvious that the sequence $w_n$ is computable. To show that $u_n$ is computable we note that the language $S_n^*$ is regular. Indeed, the set $S_n$ doesn't contain a finite set of blocks. The language $L$ consisting of all block concatenations and the empty word is regular. Thus

$$S_n^* = L \setminus \left( \bigcup_b LbL \right), \tag{2}$$

where $b$ runs over all blocks that do not belong to $S_n$. Equation (2) implies that $S_n^*$ is regular due to the well-known fact that the class of regular languages is closed under set-theoretic operations. It implies that $L(A_n')\Sigma^* \cap S_n^*$ is also regular. Note that computation of the lexicographically smallest word in a regular language is reducible to computation of the distance between vertices of a digraph.

Now we prove that the problem $R_p(W)$ is decidable. We assume that numeration of the automata is computable bijection. So there is an algorithm that computes the number $n$ of an automaton $A$. It follows from the construction that all words $u_k$, $w_k$ for $k > n$ consist of blocks with block ranks in the set $S_n$ (if a machine stops then it never starts again). The word $u_n$ enforces the transition through an accepting state if such a transition possible. Thus if the automaton $A$ doesn't pass through an accepting state after reading the prefix $v_n u_n$ of the word $W$ then it never passes through an accepting state.

On the other hand, the construction guarantees that the block $b_n$ appears infinitely often in the word $W$ iff $n$th Turing machine doesn't stop on the empty input. It implies that the problem $\mathrm{R}_f^\infty(W)$ is undecidable. Due to Proposition 1 the problem $\mathrm{R}_p^\infty(W)$ is also undecidable. □

## 3 Definitive words and languages

Now we present a new method of construction of prefix decidable and Büchi decidable infinite words. The construction is based on a notion of a definitive word. This notion is similar to the congruence relation that was introduced in Büchi paper [1] and later has been widely used in studies of decidability problems concerning $\omega$-automata.

A word $w_A \in \Sigma^*$ is called *definitive* for an automaton $A$ over the input alphabet $\Sigma$ if for any starting state either the automaton passes through an accepting state while reading the word $w_A$ or it finishes the reading in a *dead-lock* state. State $q$ is a dead-lock if there is no path from $q$ to an accepting state in the transition graph of the automaton.

Definitive words form the *definitive language* $\mathcal{D}(A)$ for the automaton $A$.

A definitive word enforces the answer in a prefix realizability problem. It means that if word $w_A$ is a factor of $W$ then the answer for an instance $A$ of the problem $\mathrm{R}_p(W)$ is determined after reading a prefix $\Sigma^* w_A$. It appears that for any automaton the definitive language is regular and non-empty.

**Proposition 3.** *There exists a definitive word for any automaton $A$.*

*Proof.* We present an algorithm producing a definitive word and show the correctness of the algorithm.

**Algorithm**. Let an automaton $A$ has $n$ states. Enumerate them. Define words $u_i$ by the following rule. If there is a path from the state $q_i$ to an accepting state $q_f$ then choose $u_i$ such that $\delta_A(u_i, q_i) = q_f$. If there is no such path then $q_i$ is a dead-lock. In this case $u_i = \varepsilon$ (the empty word).

Now define a sequence of words $w_i$, $i = 1, \ldots, n$:

$$w_1 = u_1,$$
$$w_i = w_{i-1} u_j, \qquad \text{for } i > 1, \text{where } j = \delta_A(w_{i-1}, q_i). \qquad (3)$$

The word $w_n$ is a definitive word.

**Correctness of the algorithm**. Let the automaton read the word $w_n$ starting from a state $q_i$. The word $w_i$ is a prefix of the $w_n$ by construction. It is sufficient to analyze a behavior of the automaton while reading $w_i$. We get $\delta_A(w_i, q_i) = \delta_A(u_j, q_j)$ from (3). By definition of the word $u_j$ the state $\delta_A(u_j, q_j)$ either accepting or dead-lock.

So the word $w_n$ is definitive. □

Thus the definitive language $\mathcal{D}(A)$ is non-empty. Now we show that $\mathcal{D}(A)$ is regular and give an algorithm to construct it.

**Proposition 4.** *The definitive language $\mathcal{D}(A)$ is regular.*

*Proof.* Let $A = (\Sigma, Q, \delta, q_0, F)$, where $\Sigma$ is the input alphabet,, $Q$ is the state set, $\delta$ is the transition function, $q_0$ is the initial state, $F$ is the set of accepting states. Denote by $T$ the set of dead-lock states.

For each state $q \in Q$ construct an automaton $A_q$:

$$A_q = (\Sigma, Q, \delta, q, F \cup T).$$

Let $L_q$ be a language recognizing by the automaton $A_q$. Then

$$\mathcal{D}(A) = \bigcap_{q \in Q} L_q \Sigma^*.$$

By definition of the language $\mathcal{D}(A)$ any word $w \in \mathcal{D}(A)$ belongs to $L_q \Sigma^*$ for all $q$. Indeed, there exists a prefix $w_q$ of the word $w$ such that $\delta_A(w_q, q) \in F \cup T$, so $w_q \in L_q$. Thus

$$\mathcal{D}(A) \subseteq \bigcap_{q \in Q} L_q \Sigma^*.$$

Let's prove the opposite inclusion. Take a word $w \in \bigcap_{q \in Q} L_q \Sigma^*$. If the automaton $A$ reads the word $w$ starting from an arbitrary state then the automaton either passes through an accepting state or finishes at a dead-lock state. It means that the word $w$ is definitive. □

Now we state the main results of this section.

**Theorem 2.** *The prefix realizability problem* $\mathrm{R}_p(W), W \in \Sigma^\infty$ *is decidable if any word from* $\Sigma^*$ *is a factor of the infinite word* $W$.

*Proof.* Let the input of the prefix realizability problem is a regular language $R$ recognized by a deterministic automaton $A$.

A deciding algorithm simulates an operation of the automaton $A$ on the (computable) word $W$ until either accepting or dead-lock state is encountered. The former case implies the positive answer, the latter—the negative answer.

The algorithm is correct because the infinite word $W$ has a factor $w_A \in \mathcal{D}(A)$. So, the algorithm stops. □

**Theorem 3.** *The Büchi realizability problem* $\mathrm{R}_p^\infty(W), W \in \Sigma^\infty$ *is decidable if any word from* $\Sigma^*$ *is a factor of the infinite word* $W$.

*Proof.* Let $A$ be an input of the problem $\mathrm{R}_p^\infty(W)$. The automaton $\tilde{A}$ differs from the automaton $A$ by the accepting set. Accepting states of the $\tilde{A}$ are dead-locks of the $A$.

Now we prove that $L(\tilde{A}) \cap \mathrm{Pref}(W) = \varnothing$ is equivalent to $|L(A) \cap \mathrm{Pref}(W)| = \infty$. In other words, $\mathrm{R}_p^\infty(W) \leqslant_\mathrm{m} \neg \mathrm{R}_p(W)$. Due to Theorem 2 it means that $\mathrm{R}_p^\infty(W)$ is decidable.

If $L(\tilde{A}) \cap \mathrm{Pref}(W) \neq \varnothing$ then the automaton $A$ reaches a dead-lock state while reading the word $W$. So $|L(A) \cap \mathrm{Pref}(W)| < \infty$.

In the case $L(\tilde{A}) \cap \mathrm{Pref}(W) = \varnothing$ we show that the word $W$ contains infinitely many nonintersecting factors from the language $\mathcal{D}(L(A))$. Then $|L(A) \cap \mathrm{Pref}(W)| =$

∞ because the automaton $A$ passes through either accepting or dead-lock state while reading any factor from the definitive language.

Let $w$ be a word from $\mathcal{D}(L(A))$. The word $W$ has a factor $w$. Choose an occurrence of $w$ in $W$: $w_1 = W[n_1, m_1] = w$. Construct an infinite sequence on nonintersecting factors $w$ by the following rule: if $w_k = W[n_k, m_k]$ then take an occurrence of the word $w^{2m_k}$ in the word $W$ and set $w_{k+1}$ as the suffix of this occurrence of the length $m_k - n_k + 1$. □

From Proposition 2 we have the following corollary.

**Corollary 1.** *Theory* $\mathrm{MT}(\mathbb{N}, <, W)$ *is decidable if any word from* $\Sigma^*$ *is a factor of the infinite word* $W$.

Now we present application of the above results.

**Example 1.** Consider an infinite binary word $W = 011011\ldots$ obtained by concatenation of all finite binary words taken in the lexicographic order. Theorems 2, 3 and Corollary 1 imply that the word $W$ is prefix decidable Büchi decidable and theory $\mathrm{MT}(\mathbb{N}, <, W)$ is decidable. Another proof of decidability of $\mathrm{MT}(\mathbb{N}, <, W)$ was proposed in [9].

**Example 2.** A real number $\alpha$ is called 2-*normal* if any binary word of the length $n$ appears with limiting frequency $2^{-n}$ in the binary expansion of the $\alpha$. So the binary expansion of a normal number contains any finite word as a factor. It means that the above results can be applied.

It was shown in [10] that the constant $\pi$ is 2-normal under some conjecture in dynamic system theory. So, under the same conjecture, the binary expansion of the $\pi$ is prefix decidable, Büchi decidable and the corresponding monadic theory is also decidable.

Theorem 2 can be generalized to infinite alphabets. This generalization implies more powerful condition of prefix decidability.

Let $\Sigma_\infty = \{\alpha_1, \alpha_2, \ldots, \alpha_n, \ldots\}$ be a countable alphabet and $W_\infty$ be an infinite word such that any finite word over the alphabet $\Sigma_\infty$ is a factor of $W_\infty$. It appears that for some morphisms $\varphi \colon \Sigma_\infty^* \to \Sigma^*$ the problem $\mathrm{R}_p(\varphi(W_\infty))$ is decidable.

We start from definition of finite deterministic automaton over an infinite alphabet. The definition is similar to the standard one.

**Definition 1.** A deterministic finite automaton over a countable alphabet $\Sigma_\infty$ is a 5-tuple $A_\infty(\Sigma_\infty, Q, \delta, q_0, F)$, where

- $\Sigma_\infty = \{\alpha_1, \alpha_2, \ldots, \alpha_n, \ldots\}$ is the alphabet;

- $Q$ is the finite state set;

- $\delta \colon \Sigma_\infty \times Q \to Q$ is the transition function;

- $q_0$ is the initial state;

- $F$ the set of accepting states.

Similarly to the standard definitions, a word $w \in \Sigma_\infty$ is accepted by the automaton $A$ if the automaton $A$ finishes the reading of the word $w$ at an accepting state.

The set of words accepted by the automaton $A_\infty$ is denoted by $L(A_\infty)$.

The language $L_\infty$ over the alphabet $\Sigma_\infty$ is regular if $L(A_\infty) = L_\infty$ for some automaton $A_\infty$.

The definition is obviously non-efficient. Say, the transition function $\delta$ might be noncomputable. We will consider *efficient automata* only. A class of efficient automata is defined by such a description format for automata over an infinite alphabet that the transition function $\delta$ is computable and there exists an algorithm to check the condition

$$\exists \alpha \in \Sigma_\infty : \delta(\alpha, q_1) = q_2, \quad q_1, q_2 \in Q \qquad (4)$$

having the description of the $A_\infty$.

Note that by checking the condition (4) one can compute a transition relation $\Delta \colon 2^Q \to 2^Q$, which is defined as follows

$$\Delta(A) = B \text{ iff } \quad \forall q_B \in B \, \exists q_A \in A, \, \exists \alpha \in \Sigma_\infty : \delta(\alpha, q_A) = q_B.$$

In other words, the image $\Delta(A)$ consists of those states that can be reached from the states of the set $A$ by reading a single symbol of $\Sigma_\infty$.

Our main interest is an image $L(A_\infty, \varphi)$ of language $L(A_\infty)$ recognized by the automata $A_\infty$ under a morphism $\varphi \colon \Sigma_\infty^* \to \Sigma^*$. We assume that a morphism $\varphi$ is computable.

It is worth to mention that complexity of $L(A_\infty, \varphi)$ depends on morphism $\varphi$ as well as on an automaton $A_\infty$.

**Example 3.** Let's define an automaton $A_\infty = (\Sigma_\infty, Q, \delta, q_0, F)$ by the following rule

- $Q = \{q_0, q_1\}$;
- $\delta(\alpha_{2k}, q_0) = q_0, \; \delta(\alpha_{2k+1}, q_0) = q_1, \; \delta(\alpha_k, q_1) = q_1, \; k \in \mathbb{N}$;
- $F = \{q_0\}$

and two morphisms $\varphi_1, \varphi_2 \colon \Sigma_\infty^* \to \Sigma^* = \{0, 1\}^*$ as follows

$$\varphi_1(\alpha_{2k}) = 0^k 1^k, \; \varphi_1(\alpha_{2k+1}) = 1^k,$$
$$\varphi_2(\alpha_{2k}) = 0^k, \; \varphi_2(\alpha_{2k+1}) = 1^k.$$

Then $L(A_\infty, \varphi_1) = \{0^k 1^k\}$ is a CFL while $L(A_\infty, \varphi_2) = \{0^k\}$ is regular.

Note that the language $L(A_\infty, \varphi)$ is regular if for any pair of states $q_i, q_j \in Q$ languages $R_{i,j} = \{\varphi(\alpha_k) \mid \delta(\alpha_k, q_i) = q_j\}$ are regular.

To simplify notations we introduce problems $\mathrm{R}_p(W_\infty, \varphi)$ as $\mathrm{R}_p(\varphi(W_\infty))$.

We are going to find conditions on a morphism $\varphi$ that guarantee prefix decidability of the problem $\mathrm{R}_p(W_\infty, \varphi)$ provided that any word in $\Sigma_\infty^*$ is a factor of $W_\infty$. To this aim we generalize Theorem 2 to countable alphabets and then construct a reduction of $\mathrm{R}_p(W_\infty, \varphi)$ to $\mathrm{R}_p(W_\infty)$.

Definitive words for the infinite alphabet case are defined in the same way as for the finite alphabet. It is easy to see that Proposition 3 on an existence of a definitive word also holds for inifinite alphabets because the proof do not depend on the size of the alphabet. A generalization of Theorem 2 holds for efficient automata.

**Proposition 5.** *The problem $R_p(W_\infty)$ is decidable for efficient automata provided any word from $\Sigma_\infty^*$ is a factor of $W_\infty$.*

*Proof.* At the first stage of an algorithm one should check for each state $q$ an existence of a path from the state $q$ to the set of accepting states. Equivalently, the orbit $\Delta^n(\{q\})$ has nonempty intersection with the set $F$ of accepting states, i.e. $\exists n \in \mathbb{N} : \Delta^n(\{q\}) \cap F \neq \varnothing$. The state set is finite. So the orbit is also finite. For an efficient automaton it can be found algorithmically.

If there is no such path from the initial state the answer is negative.

Otherwise, mark as dead-locks all states such that there is no path to the set $F$. Then simulate an operation of the automaton $A_\infty$ while reading $W_\infty$. A definitive word is a factor of $W_\infty$. So, at some moment of time the automaton state will be either accepting or a dead-lock. □

Now we state the sufficient conditions for an existence of a reduction $R_p(W_\infty, \varphi)$ to $R_p(W_\infty)$. The basic condition is a decidability of regular realizability problem for the language $L_\varphi$ consisting of images of symbols of the alphabet $\Sigma_\infty$.

**Proposition 6.** *Suppose that a map $\alpha \mapsto \varphi(\alpha)$ is computable and for the language $L_\varphi = \{\varphi(\alpha) \mid \alpha \in \Sigma_\infty\}$ the regular realizability problem is decidable.*

*Then there exists an algorithm that takes an input $A$ of the prefix realizability problem $R_p(W_\infty, \varphi)$ and outputs an automaton $A_\infty$ such that the answer in the problem $R_p(W_\infty, \varphi)$ for the language $L(A)$ is the same as the answer in the problem $R_p(W_\infty)$ for the language $L(A_\infty)$. The outputs of the algorithm form a class of efficient automata.*

*Proof.* Take an automaton $A = (\Sigma, Q, \delta, q_0, F)$. It will be reduced to an automaton $A_\infty = (\Sigma_\infty, \tilde{Q}, \tilde{\delta}, \tilde{q}_0, \tilde{F})$. Here $\tilde{Q} = Q \times \{0, 1\}$. An auxiliary bit will be used to store the fact that the automaton $A$ passed through an accepting state along a path to the state $q$. An element $(q, b) \in \tilde{Q}$ will be denoted by $\tilde{Q}$ if the value of the auxiliary bit is irrelevent.

Replacing the initial state of the $A$ by a state $q_i$ and the accepting set by the set $\{q_j\}$ we obtain an automaton recognizing a regular language $R_{i,j}$ consisting of all paths from the state $q_i$ to the state $q_j$. We use a notation $R_{i,j}$ as a regular expression. So $R_{i,k}R_{k,j}$ is also a regular expression for those paths from $q_i$ to $q_j$ that pass through $q_k$.

The transition function $\tilde{\delta}$ of the automaton $A_\infty$ is defined as follows

$$\begin{aligned} \tilde{\delta}(\alpha, \tilde{q}_i) = (q_j, 1), \ \exists q_k \in F : \varphi(\alpha) \in R_{i,k}R_{k,j}, \\ \tilde{\delta}(\alpha, \tilde{q}_i) = (q_j, 0), \ \varphi(\alpha) \in R_{i,j}, \forall q_k \in F : \varphi(\alpha) \notin R_{i,k}R_{k,j}. \end{aligned} \quad (5)$$

Equations (5) mean that there is an $\alpha$-transition in $A_\infty$ from a state $(q_i, a)$ to a state $(q_j, b)$ iff there is a transition in $A$ from the state $q_i$ to the state $q_j$ by reading the word $\varphi(\alpha)$. Additionally, if the automaton $A$ passes through an accepting state during the transition $q_i \xrightarrow{\varphi(\alpha)} q_j$ then the auxiliary bit $b = 1$, otherwise $b = 0$.

Define the set of accepting states of the automaton $A_\infty$ as the set of all states with the value 1 of the auxiliary bit, in other words $\tilde{F} = (q, 1) \in \tilde{Q}$. The initial state of the $A_\infty$ is $(q_0, 0)$.

Note that the automata $A_\infty$ form a class of efficient automata. The transition function (5) is computable because it's value is determined by regular events. The condition (4) is verified efficiently because the check for a transition $\tilde{q}_i \xrightarrow{\Sigma_\infty} \tilde{q}_j$ is reduced to a number of checks of the form $R_{i,k} R_{k,j} \cap L_\varphi \neq \varnothing$. The latter can be done because the regular realizability problem $\mathrm{RR}(L_\varphi)$ is decidable.

Now we prove the correctness of the above reduction.

Suppose that $L(A) \cap \varphi(W_\infty) \neq \varnothing$. It means that $A$ passes through an accepting state while reading the infinite word $\varphi(W_\infty)$. Let $m \in \mathbb{N}$ be the smallest positive integer such that the automaton $A$ passes through an accepting state while reading the image of the prefix $\varphi(W_\infty[1, m])$. Look at operation of the automaton $A_\infty$ on $W_\infty[1, m]$. By definition we have

$$\tilde{\delta}(W_\infty[i], \tilde{q}_j) = (\delta(\varphi(W_\infty[i]), q_j), b),$$

where $b$ is the auxiliary bit. Thus

$$\tilde{\delta}(W_\infty[1, m], \tilde{q}_0) = (\delta(\varphi(W_\infty[1, m]), q_0), 1).$$

In the opposite direction the proof is basically the same. Suppose that after reading a prefix $W_\infty[1, m]$ the automaton $A_\infty$ is at an accepting state, i.e. the auxiliary bit is set to 1. It means that while reading $\varphi(W_\infty[1, m])$ the automaton $A$ passes through an accepting state. □

Propositions 5 and 6 imply the following theorem.

**Theorem 4.** *Suppose that map $\alpha \mapsto \varphi(\alpha)$ is computable and for the language $L_\varphi = \{\varphi(\alpha) \mid \alpha \in \Sigma_\infty\}$ the L-realizability problem is decidable.*

*If any word of $\Sigma_\infty^*$ is a factor of an infinite word $W_\infty$ then*
*– the prefix realizability problem $\mathrm{R}_p(W_\infty, \varphi)$ is decidable;*
*– the Büchi realizability problem $\mathrm{R}_p^\infty(\varphi(W_\infty))$ is decidable;*
*– the theory $\mathrm{MT}(\mathbb{N}, <, W)$ is decidable.*

The first statement of the thorem is proved above. The second and the third are proved in the same way as Theorem 3 and Corollary 1. Indeed, the arguments in the proof of Theorem 3 do not depend on the size of the alphabet.

## 4 An equivalence of regular realizability and prefix realizability

In this section we apply the above results to show an equivalence of decidability for regular realizability and prefix realizability problems.

For this purpose we will use an infinite alphabet to construct a prefix decidable word $W_\infty$. Then we choose an appropriate morphism $\varphi$ such that a problem $\mathrm{RR}(L)$ is reduced to the problem $\mathrm{R}_p(W_\infty, \varphi)$.

We will use the following alphabets

$\Sigma = \{0, 1\}$;

$\Sigma_\# = \{0, 1, \#\}$;

$\Sigma_\infty = \{\alpha_1, \alpha_2, \alpha_3, \ldots, \alpha_n, \ldots\}$ is an infinite alphabet.

Let $L$ be a language over the binary alphabet $\Sigma$ such that $\mathrm{RR}(L)$ is decidable. Hence the language $L$ is decidable and recursively enumerable. Fix an enumeration of $L$ and denote $i$th word by $w_i$. Define a morphism $\varphi\colon \Sigma_\infty \to \Sigma_\#$ such that $\varphi(\alpha_i) = w_i\#$ and an infinite word $W$ as $\varphi(W_\infty)$.

Choose an infinite word $W_\infty$ over the alphabet $\Sigma_\infty$ such that any word in $\Sigma_\infty^*$ is a factor of $W_\infty$. Then Theorem 4 implies decidability of the problem $\mathrm{R}_p(W_\infty, \varphi) = \mathrm{R}_p(W)$ provided the problem $\mathrm{RR}(L)$ is decidable. (Actually, we need decidability of the problem $\mathrm{RR}(L_\varphi) = \mathrm{RR}(L\#)$ but this problem is equivalent to the $\mathrm{RR}(L)$). It follows that

$$\mathrm{R}_p(W) \leqslant_\mathrm{T} \mathrm{RR}(L),$$

where $\leqslant_\mathrm{T}$ is a Turing reducibility.

In the opposite direction, let $R$ be an input of the problem $\mathrm{RR}(L)$. Define a language $\tilde{R} = (\Sigma^*\#)^* R\#$. The condition $\tilde{R} \cap W \neq \varnothing$ is equivalent to the condition $R \cap L \neq \varnothing$. So

$$\mathrm{RR}(L) \leqslant_\mathrm{m} \mathrm{R}_p(W).$$

# References

[1] J. R. Büchi. On a decision method in restricted second-order arithmetic. In Proceedings of International Congress for Logic, Methodology and Philosophy of Science, pages 1–11. Stanford University Press, 1962.

[2] A. L. Semënov. Logical theories of one-place functions on the set of natural numbers. Mathematics of the USSR – Izvestiya (1984), 22(3):587–618.

[3] Carton O., Thomas W. The Monadic Theory of Morphic Infinite Words and Generalizations. Information and Computation Volume 176, Issue 1, 10 July 2002. Pages 51–65.

[4] *Vyalyi M.N.* On models of a nondeterministic computation Proc. of CSR 2009. Springer Lecture Notes in Computer Science, vol. 5675, 2009. P. 334–345.

[5] M.N. Vyalyi. On models of a nondeterministic computation for 2-sided automata. Proc. of VIII international conference "Discrete models in control system theory". Moscow, MaxPress, 2009. P. 54–60. (in Russian)

[6] A.A. Muchnik, Y.L. Pritykin, A.L. Semenov. Sequences close to periodic. Russian Mathematical Surveys, 2009, 64:5, 805–871.

[7] R. McNaughton. Testing and generating infinite sequences by a finite automaton. Information and Control, 9:521–530, 1966.

[8] M. Roggenbach. Determinization of Büchi-Automata. In E.Grädel, W.Thomas, T.Wilke (eds): Automata, Logics, and Infinite Games, LNCS 2500, Springer 2002. P. 43–60.

[9] Bárány V. A Hierarchy of Automatic $\omega$-Words having a Decidable MSO Theory. Informatique Théorique et Applications. Volume 42, 2008. P. 417–450.

[10] David H. Bailey and Richard E. Crandall. On the random character of fundamental constant expansions Experimental Mathematics, 2001. Vol. 10, no 2. P. 175–190.